\documentclass[conference]{IEEEtran}
\IEEEoverridecommandlockouts
\usepackage{cite}
\usepackage{amsmath,amssymb,amsfonts}
\usepackage{algorithmic}
\usepackage{graphicx}
\usepackage{textcomp}
\usepackage{xcolor}
\usepackage{booktabs}
\usepackage{array}
\usepackage{multirow}
\usepackage{url}
\newcommand{\system}{ULTRAS}

\newcommand{\systemexpbold}{\textbf{U}nified \textbf{L}earning of \textbf{T}ransformer \textbf{R}epresentations for \textbf{A}udio and \textbf{S}peech}

\def\BibTeX{{\rm B\kern-.05em{\sc i\kern-.025em b}\kern-.08em
    T\kern-.1667em\lower.7ex\hbox{E}\kern-.125emX}}
\begin{document}

\title{ULTRAS - Unified Learning of Transformer Representations for Audio and Speech Signals\\
}
\author{Ameenudeen P E, Charumathi Narayanan,  and Sriram Ganapathy \\ 
\IEEEauthorblockA{LEAP Laboratory, Electrical Engineering, Indian Institute of Science, Bangalore, India.}
\texttt{ameenudeenp@iisc.ac.in}\thanks{This work was performed with grants received from   Ministry of Information Technology (MEiTY) NLTM, India. }
 }

\maketitle

\begin{abstract}
Self-supervised learning (SSL) has driven impressive advances in speech processing by adopting time-domain prediction objectives, while audio representation learning frameworks operate on time-frequency spectrograms. Models optimized for one paradigm  struggle to transfer to the other, highlighting the need for a  joint framework. 
We propose  \systemexpbold{} (\system), where the masking and predictive modeling is performed over long patches of the data. 
The model, based on the transformer architecture, encodes spectral-patches of log-mel spectrogram features. The predictive modeling of masked segments is performed on spectral  and temporal targets using a combined loss-function, forcing the representations to encode time and frequency traits.  
Experiments are performed on a variety of speech and audio tasks, where we illustrate that the \system{} framework achieves improved performance over other established baselines. 
\end{abstract}


\section{Introduction}
Self-supervised learning (SSL) has revolutionized representation learning across various input modalities. In natural language processing, models like BERT \cite{devlin2019bert}, introduced masked language modeling (MLM) on deep bidirectional transformers, achieving state-of-the-art results on many tasks. In computer vision, masked auto-encoding (MAE) has proven similarly effective: for example, He \emph{et al.} \cite{he2022masked} reports that randomly masking a significant proportion of image patches and reconstructing them (MAE framework) yields scalable learners. The Vision-Transformer \cite{dosovitskiy2020image} has delivered performance breakthroughs when coupled with the SSL frameworks. 
In the audio and speech domain, SSL methods have also made significant inroads. Models such as wav2vec 2.0 \cite{baevski2020wav2vec} and HuBERT \cite{hsu2021hubert} apply masked prediction on raw waveform or feature sequences, learning powerful encoders of 1-D temporal representations of speech. These approaches exploit sequential structure in speech and are well suited for tasks like automatic speech recognition (ASR), emotion recognition and speaker recognition.

On the other hand, general audio signals (e.g.\ environmental sounds, music, etc.)  often carry information in the 2-D spectro-temporal domain. 
The standard convolutional networks  or spectrogram-transformers ingest patches of log-mel spectrograms and model the spectral patterns\cite{gong2021ast}. For instance, Gong \emph{et al.} \cite{gong2022ssast} demonstrates that a spectrogram-based transformer (SSAST) with masked spectrogram-patch modeling dramatically improves performance on audio event classification tasks. 
However, these approaches fail to generalize to speech related tasks. 

Beyond uni-modal audio, several recent multi-modal frameworks incorporate audio representations. Zhu \emph{et al.} \cite{zhu2023vatlm} propose VATLM, which unifies visual, audio, and text inputs into a shared semantic space via a masked unified-token prediction objective. Choi \emph{et al.} \cite{choi2024av2av} introduce AV2AV, a system for audio-visual speech translation that learns modality-agnostic speech representations (leveraging AV-HuBERT) so that a single model can translate speech without intermediate text. Likewise, Su \emph{et al.} \cite{su2024vision} introduced VAB, which encodes video frames via a pre-trained image encoder and audio via a neural codec, and applied masked audio-token prediction conditioned on visual context to learn a joint audio-visual model. These multimodal SSL models demonstrate the feasibility of aligning audio with other modalities, but require the presence of multiple modalities in the input data.

There are limited efforts on attempting to jointly model speech and naturalistic audio signals using the same SSL framework.
BEATs encoder  \cite{chen2022beats} introduces an acoustic tokenizer and an iterative discrete-label prediction objective, bringing the textual discrete SSL paradigm to general audio data.
The EnCodecMAE \cite{pepino2023encodecmae} attempts to perform universal modeling using a masked autoencoder and a neural audio codec. 
A teacher-student framework for clip-level and frame-level representation learning was proposed by Xian et al.~\cite{li2024self}. 
In another work, Nilzumi et al.~\cite{niizumi2024masked} proposed the learning of separate models for masked and unmasked regions of the audio. 
However, none of these frameworks explicitly model the joint temporal and spectral information present in the audio signal, leading to a degradation in the diverse evaluation of speech and audio tasks. 




\begin{figure*}[t!]
    \centering
    \includegraphics[width=0.7\linewidth]{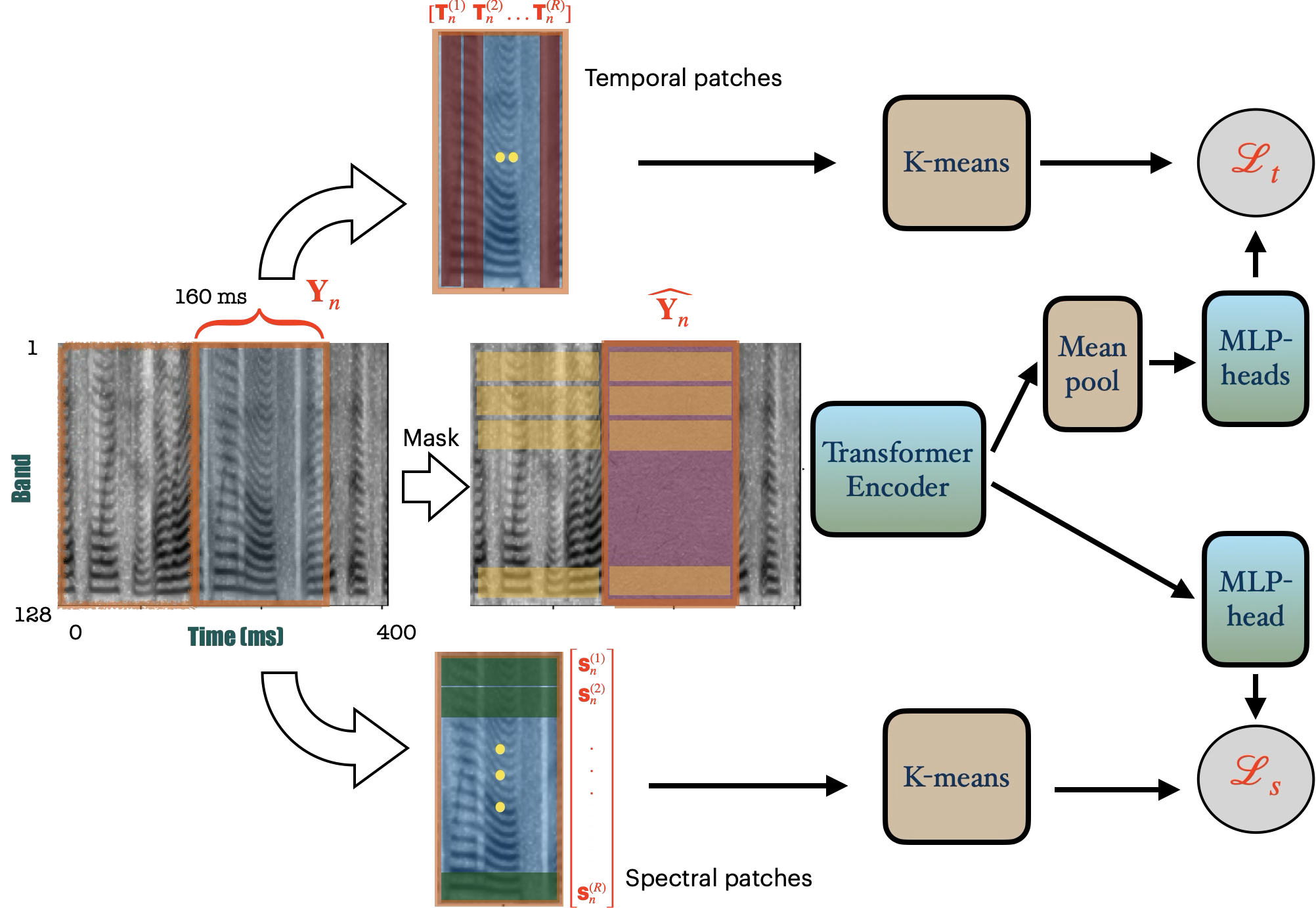}
    \caption{Block schematic of the proposed framework of joint 1-D and 2-D modeling of audio data. The gradient colored blocks are learnable, while the rest do not have any learnable parameters.}
    \label{fig:block-schematic}
    \vspace{-0.1in}

\end{figure*}

In this paper, we propose \systemexpbold{} (\system), an approach to jointly model the time-frequency attributes of the input acoustic signal.  
Unlike the conventional speech SSL approaches that encode units of short windows ($20$ms), the proposed framework masks relatively long duration audio segments ($160$ms windows).
The \system{} 
encodes spectral patches of the input signal using a stack of transformer layers. The collection of spectral patches belonging to the same segment allows the predictive modeling of both spectral codes (discrete symbol representations of spectral patches) as well as the temporal codes (discrete symbol representation of temporal frames). 
The joint predictive task encourages the representations to embed spectral and temporal language modeling traits, which are important for downstream tasks. 

The proposed model is pre-trained on a combined dataset of speech and audio signals. The evaluations are performed on various downstream tasks, where the SSL model is frozen and only a light-weight classification head is trained for the supervised task. 
We compare the proposed SSL with several other established SSL frameworks on utterance-level tasks in speech, music and audio-event related domains. In these tasks, the \system{} illustrates improved performance highlighting the benefits of the joint spectro-temporal predictive tasks.  


The key contributions of this work are:-
\begin{itemize}
\item Masked modeling of  long audio segments of syllable length for effective encoding of acoustic information. 
\item Joint prediction of spectral and temporal targets for embedding time-frequency characteristics. 
\item Comprehensive evaluation on a diverse set of speech and audio downstream tasks to illustrate the effectiveness of the work.

\end{itemize}

\section{Proposed \system{} Framework}
Figure~\ref{fig:block-schematic} illustrates the main workflow of our proposed iterative audio pre-training framework. Our approach leverages a Vision Transformer-style SSL model that takes a 2-D audio spectrogram as input.  The model is optimized through an unified 
loss function. 
The goal is to learn robust and generalizable audio representations applicable to both speech and  audio tasks.
\subsection{Input pre-processing}
The input audio waveform is first transformed into a log-mel spectrogram representation. Specifically, we compute $128$-dimensional mel-filterbank (fbank) features using a $25$ ms Hanning window with a $10$ ms hop size.
Let the spectrogram be denoted as  $\boldsymbol{\mathcal{X}} = [\textbf{x}_1, \textbf{x}_2, \dots, \textbf{x}_M]$, 
where $M$ is the number of time-frames  and $\textbf{x} _ t \in \mathcal{R} ^{D}$, $D=128$.  
\subsection{Windowing and modeling}
The spectrogram is divided into non-overlapping windows of $P$ frames. In our experiments, we use $P=16$ frames, corresponding to $160$ms of audio. Let the windowed  input be denoted as 
$\boldsymbol{\mathcal{Y}} = [\textbf{Y}_1, \textbf{Y}_2, \dots, \textbf{Y}_N]$. Here, $\textbf{Y}_n$ is a windowed spectrogram of size $\mathcal{R}^{D \times P}$, where $N=M/P$
\subsection{Random Masking Strategy}
To enable self-supervised learning, we apply a random masking strategy to the windowed spectrograms $\boldsymbol{\mathcal{Y}}$. 
Each segment is  masked with a probability $p$. Additionally, to encourage the model to learn long contextual dependencies, if a patch is masked, its subsequent patch is also masked with a fixed probability of $p'$.

Let the resulting masked sequence be denoted as:
\[
\boldsymbol{\widehat{\mathcal{Y}}} = [\widehat{\textbf{Y}}_1, \widehat{\textbf{Y}}_2, \dots, \widehat{\textbf{Y}}_N]
\]
where:
\[
\widehat{\textbf{Y}}_n = 
\begin{cases}
\texttt{[MASK]}, & \text{if } \text{location } $n$ \text{ is selected for masking} \\
\textbf{Y}_n, & \text{otherwise}
\end{cases}
\]
In our experiments, $p=0.6$ and $p'=0.2$.

\subsection{Spectral Embedding and Transformer Encoding}
Each windowed spectrogram $\textbf{Y}_n \in \mathcal{R}^{D \times P}$ is uniformly partitioned into $R$ non-overlapping spectral patches, each of size $P \times P$. These patches are denoted as:
\[
\textbf{Y}_n = \begin{bmatrix}
  \textbf{S}_n^{(1)} \\
  \textbf{S}_n^{(2)} \\
  \dots \\
  \textbf{S}_n^{(R)}
\end{bmatrix} ;  \quad \textbf{S}_n^{(i)} \in \mathcal{R}^{P \times P},
\] where $R=\dfrac{D}{P}$

Each spectral patch $\textbf{S}_n^{(i)}$ is flattened into a vector and added with a learnable positional embedding to retain spatial context. The resulting sequence of vectors, obtained from patched spectrogram inputs, is passed through the transformer encoder, which outputs a sequence of embeddings:
\[
\textbf{Z}_n = [\textbf{z}_n^{(1)}, \textbf{z}_n^{(2)}, \dots, \textbf{z}_n^{(R)}], \quad \textbf{z}_n^{(i)} \in \mathcal{R}^{D' \times 1}
\]
With the masked input $\boldsymbol{\widehat{\mathcal{Y}}}$ to the transformer, the outputs are, 
\[
\widehat{\boldsymbol{\mathcal{Z}}} = [\widehat{\textbf{Z}}_1, \widehat{\textbf{Z}}_2, \dots, \widehat{\textbf{Z}}_N],
\]
where each $\widehat{\textbf{Z}}_n$ consists of $R$ encoded patch embeddings of the masked input.

\subsection{Spectral Targets and Loss Function}

Each spectrogram patch $\textbf{S}_n^{(k)} \in \mathbb{R}^{P \times P}$ (for $k = 1, \dots, R$) is flattened and quantized using a  K-means quantizer $\mathcal{Q}$ to obtain discrete targets representing frequency bin patterns. These targets are denoted as:
\[
\textbf{C}_{s,n}^{(k)} = \mathcal{Q}(\textbf{S}_n^{(k)}), \quad \textbf{C}_{s,n}^{(k)} \in \{1, 2, \dots, K_s\}
\]
where $K_s$ is the number of spectral-patch clusters (codebook size) in the quantizer.

The model outputs corresponding to the masked patches $\widehat{\textbf{z}}_n^{(k)} \in \mathbb{R}^{D' \times 1}$ are projected using a learnable multi-layer perceptron (MLP) head, $\textbf{W}_s \in \mathbb{R}^{K_s \times D'}$, followed by a softmax activation to produce cluster probabilities:
\[
\widehat{\textbf{P}}_{n}^{(k)} = \text{softmax}(\textbf{W}_s \widehat{\textbf{Z}}_n^{(k)})
\]

The spectral-loss function is the cross-entropy loss between the predicted distribution and the quantized target label for each masked patch:

\begin{equation}\label{eqn:spectral-loss}
\mathcal{L}_{\texttt{s}} = \frac{1}{|\mathcal{M}|\cdot R} \sum_{n \in \mathcal{M}} \sum_{k=1}^R\text{C.E.}  \left( \widehat{\textbf{P}}_{n}^{(k)}, \textbf{C}_{s,n}^{(k)} \right)
\end{equation}

where $\mathcal{M}$ denotes the set of masked window indices, and $\text{C.E.}(\cdot, \cdot)$ denotes the standard cross-entropy loss.
\subsection{Temporal Targets and Loss}
Each $\textbf{Y}_n \in \mathbb{R}^{D \times P}$ can also be partitioned temporally into $R$ non-overlapping frames, i.e., 
\[
\textbf{Y}_n = [\textbf{T}_n^{(1)} ~ \textbf{T}_n^{(2)}~  \dots ~ \textbf{T}_n^{(R)}], \quad \textbf{T}_n^{(j)} \in \mathcal{R}^{D \times {P'}}
\]
 
The temporal frames $\textbf{T}_n^{(j)}$,  are  quantized using a K-means quantizer $\mathcal{Q}$ trained with $K_t$ clusters, resulting in discrete frame-level targets:
\[
\mathcal{Q}(\textbf{T}_{n}^{(k)}) = \textbf{C}_{t,n}^{(k)} \quad \text{where } \textbf{C}_{t,n}^{(k)} \in \{1, 2, \dots, K_t\}
\]
These targets represent the temporal information   across all the spectral patches. Hence, the spectral embeddings, $\widehat{\textbf{z}}_n^i $, are mean-pooled, 
\[
\widehat{\textbf{z}}_n = \frac{1}{R} \sum_{k=1}^{R} \widehat{\textbf{z}}_n^{(k)}.
\]
The temporal softmax prediction for the frame is given by:
\[
\widehat{\textbf{P}}_{n}^{(j)} = \text{softmax}(\textbf{W}_t^{j} \widehat{\textbf{z}}_n),
\]
where $\textbf{W}_t^{j}$ are learnable parameters of the MLP classification head. 

The training objective is the average cross-entropy loss over all $R$ frames of each masked window:
\begin{equation}\label{eqn:temporal-loss}
\mathcal{L}_{\texttt{t}} = \frac{1}{|\mathcal{M}| \cdot R} \sum_{n \in \mathcal{M}} \sum_{j=1}^{R} \text{C.E.} \left( \widehat{\textbf{P}}_{n}^{(j)}, \textbf{C}_{t,n}^{(j)} \right)
\end{equation}

\subsection{Total Loss}
The overall training objective of the \system{}  is a weighted combination of the spectral loss $\mathcal{L}_{\text{s}}$ and  time-frame loss $\mathcal{L}_{\text{t}}$:
\begin{equation}\label{eqn:total-loss}
\mathcal{L}_{\text{total}} = \lambda \mathcal{L}_{\texttt{t}} + (1 - \lambda) \mathcal{L}_{\texttt{s}}
\end{equation}
where $\lambda \in [0, 1]$ is a tunable hyperparameter to balance the two components.

\subsection{Implementation Details}
The proposed \system{} framework, as shown in Figure~\ref{fig:block-schematic}, consists of a spectral target as well as a temporal target. 
Further, the input spectrograms are masked and passed through a spectral encoder (stack of transformer layers) to embed the spectrogram patches. 
In our implementation, we train the \system{} with both speech and audio inputs, represented as mel-spectrogram features. \\\\
\noindent \textbf{Inputs}: Each input audio recording is truncated or zero-padded to a fixed duration of $8$ seconds. The waveform is then transformed into log-Mel spectrogram using a $25$\,ms Hanning window with a $10$\,ms hop size, producing a spectrogram of shape $\mathcal{X} \in \mathbb{R}^{D \times M}$, with $D=128$ and $M=800$. 

The spectrogram is partitioned into blocks of $P$ temporal frames, and further split into $P\times P$ patches, with $P=16$. Hence, each spectral patch ($\textbf{S}_n^{(i)}$) represents time-frequency information corresponding to $160$ms segments of the input audio and for $16$ mel-spectral bands. 
For a given spectrogram block of $160$ms, this would result in $R=8$ spectral patches. 
\\

\noindent \textbf{Transformer Encoder}: Each patch $\textbf{S}_n^{(i)}$ is vectorized and linearly projected to generate a $D' = 768$-dimensional patch embedding. These embeddings are combined with learnable positional embeddings and input to the Transformer encoder.
The architecture of the transformer encoder is similar to the Vision-Transformer (ViT)~\cite{dosovitskiy2020image}, consisting of $12$ layers with $12$ attention heads.\\

\noindent \textbf{Targets}:
For the spectral K-means clustering,  the spectrogram patches of size $P \times P$ patches are vectorized and projected to $256$-dimensional space. The k-means clustering is performed with a codebook size of $K_s = 100$ clusters and with Euclidean distance metric. 
For the temporal K-means clustering, we use spectrogram partitions of size $D \times P'$, with $P'=2$ in our case. This would mean the encoding of $20$ms chunks of audio. 
While the spectrograms could be directly quantized, we use a pre-trained HuBERT encoder~\cite{hsu2021hubert} for generating embeddings at $50$Hz sampling rate from the raw waveform. These embeddings from the $6$-th HuBERT layer are of $768$ dimensional and are vector quantized to $K_t = 500$ clusters to generate the temporal targets. 
\\

\noindent \textbf{Initialization}: Before the joint training with the spectral and temporal targets, we pre-train the \system{} model only with the spectral targets. For this training, we use a masking strategy inspired by SSAST~\cite{gong2022ssast}. Specifically, we randomly mask about $60$\% of the spectrogram patches of size $P \times P$ and train the model using the masked language modeling (MLM) loss. 
As shown in Figure~\ref{fig:block-schematic}, this would pre-train the transformer encoder and the MLP head corresponding to the spectral loss. 
This pre-training is performed for $100$k steps, before the joint training with the combined loss function (Eq:~\ref{eqn:total-loss}). 
\\

\noindent \textbf{Model Training}: 
Following the initialization, we continue the training using joint spectro-temporal masking and predictive modeling with the combined loss function. The total number of pre-training steps is set to $150k$. We use the AdamW optimizer with weight decay of $0.05$ and $\beta=[0.9,0.98]$. A linear learning rate scheduler is applied, where the learning rate linearly increases from $1e-6$ to $1e-4$ during the first $10$\% of training steps (warm-up) and then decays linearly for the remaining steps to the final value of $1e-6$.

\section{Experimental Setup}
We train the proposed \system{} on a balanced dataset comprising an equal number of speech and audio recordings, sourced from the LibriSpeech corpus~\cite{panayotov2015librispeech}  and the AudioSet corpus~\cite{gemmeke2017audio}, respectively. 
\subsection{Pre-training Data}
\noindent \textbf{$\textbf{200}$ hour setup}: 
We pre-train the proposed \system{} as well as other baseline SSL frameworks  on a $200$-hour dataset comprising of $100$ hours each from Librispeech and AudioSet dataset. 
With this small setup, we also perform several ablation experiment,  which are reported in Section IV. \\
\\
\noindent \textbf{$\textbf{2000}$ hour setup}: In this setting, we use the entire Librispeech ($1000$ hours)\cite{panayotov2015librispeech}  and a random $1000$ hour partition of the AudioSet data\cite{gemmeke2017audio}. 
These experiments were designed to identify the scalability of the proposed setting. We compare this framework with various other published works on the downstream tasks considered.

\subsection{Evaluation Setting}
The effectiveness of the learned representations is evaluated across six downstream tasks—three from the speech domain and three from the general audio domain.
For downstream evaluations, we adopt the   evaluation protocol from SUPERB (Speech Processing Universal PERformance Benchmark) \cite{yang2021superb}.
In particular, all the evaluations use the pre-trained self-supervised model as a frozen embedding extractor. For each downstream task evaluation,  a weighted sum of the hidden representations from each layer of the SSL encoder is passed to a light-weight task-specific prediction head.
During evaluation, only the parameters of the layer-wise aggregation and the prediction head for the given task are updated, while the rest of the model remains frozen.
This way of evaluation measures the quality of the pre-training representations directly without the influence of the task specific fine-tuning.

\begin{table*}[t!]
\caption{Accuracy (\%) of Different Models Pretrained on $200$ hour and $2000$ hour pre-training setup, where all the systems use the same training data and number of pre-training steps. The models also have similar parameter size.  }
\begin{center}
\resizebox{0.63\linewidth}{!}{ 
\begin{tabular}{|l|c|c|c||c|c|c|}
\midrule 
\midrule 
\multicolumn{7}{|c|}{200 hour setup} \\  \midrule 
\multirow{2}{*}{\textbf{Model}} & \multicolumn{3}{c||}{Audio}  & \multicolumn{3}{c|}{Speech}  \\ \cmidrule{2-7}  
& \textbf{\textit{ESC-50}} & \textbf{\textit{US8k}} &  \textbf{\textit{NSYNTH}} & \textbf{\textit{IEMOCAP}} & \textbf{\textit{VOX1}}  & \textbf{\textit{SPCV2}} \\
\midrule 

SSAST \cite{gong2022ssast} & 56.75 & 65.54 & 70.78 & 52.56 & 14.58 &  61.22 \\

HuBERT \cite{hsu2021hubert} & 66.10 & 70.05 & 64.43 & 59.91 & 28.83 &  84.82 \\
\midrule 
SSAST~\cite{gong2022ssast}+MLM & 77.65 & 78.46 & 74.32  & 60.49 & 32.19 &  78.89 \\

\system & \textbf{86.00} & \textbf{84.12}  & \textbf{75.83} & \textbf{63.82}  & \textbf{46.47}  &  \textbf{91.95}  \\
\midrule 
\midrule 
\multicolumn{7}{|c|}{2000 hour setup} \\ 
\midrule 
\midrule 
SSAST \cite{gong2022ssast} & 63.40 & 70.10 & 73.21 & 54.79 & 16.94 &  74.95 \\
HuBERT \cite{hsu2021hubert} & 80.55 & 78.42 & 68.23 & 66.27 & 64.56 &  \textbf{95.55} \\
\midrule 
SSAST~\cite{gong2022ssast}+MLM & 85.50& 84.27 & \textbf{76.57} & 62.05 & 47.95 &  82.03 \\

\system &  \textbf{91.15} & \textbf{86.07} & 76.52 & \textbf{67.78} &  \textbf{73.55} & 95.10 \\
\midrule 
\midrule 
\multicolumn{7}{l}{$^{\mathrm{b}}$Highest accuracy for each task is shown in bold.}
\end{tabular}}
\vspace{-0.1in}
\label{tab:downstream_results_ieee}
\end{center}
\end{table*}

\subsection{Downstream Tasks}

We evaluate  on six downstream classification tasks, spanning a variety of domains: environmental sound, speech, and music. These have been used as benchmark tasks in various prior works \cite{hsu2021hubert,yang2021superb, chen2022beats}. 
\begin{table*}[t!]
\caption{Comparison of the proposed \system{} with other baseline systems. In these settings, we have used the pre-trained checkpoint available 
 online and performed downstream evaluation by fine-tuning the light-weight head. Except for the HuBERT-base model, all other works use significantly higher amount of pre-training data compared to \system{}.}
\begin{center}
\resizebox{0.87\linewidth}{!}{ 
\begin{tabular}{|l|c|c|c|c|c||c|c|c|}
\midrule  
\midrule 
\multirow{2}{*}{\textbf{Model}} & \multirow{2}{*}{\textbf{\#Param. (M)}} & \multirow{2}{*}{\textbf{Data (hrs)}} & \multicolumn{3}{c||}{\textbf{Audio}}  & \multicolumn{3}{c|}{\textbf{Speech}}  \\ \cmidrule{4-9}
& & & \textbf{\textit{ESC-50}} & \textbf{\textit{US8k}} &  \textbf{\textit{NSYNTH}} & \textbf{\textit{IEMOCAP}} & \textbf{\textit{VOX1}}  & \textbf{\textit{SPCV2}} \\ \midrule 
HuBERT \cite{hsu2021hubert} & 95 & 960 & 77.45 & 77.61 & 69.40 & \textbf{68.30} & \textbf{86.68} & 96.10 \\
SSAST \cite{gong2022ssast} & 89 & 5,800 & 37.85 & 54.87 & 60.78 & 55.21 & 15.62 & 37.87 \\
BEATs \cite{chen2022beats} & 91 & 5,800 & \textbf{92.45} & 83.21 & \textbf{79.14} & 67.61 & 56.84 & 92.30 \\

EnCodecMAE \cite{pepino2023encodecmae} & 86.6 & 11,000 & 88.25 & 85.42 & 77.23 & 67.80 & 71.65 & \textbf{97.30} \\
\midrule 
\system & 89 & 2,000 & 91.15 & \textbf{86.07} & 76.52 & 67.78 & 73.55 & 95.10 \\
\midrule
\midrule 
\end{tabular}
}
\end{center}
\label{tab:downstream_results_ieee_2000h}
\vspace{-0.2in} 
\end{table*}
\begin{itemize}
    \item \textbf{ESC-50} \cite{piczak2015esc}: A single-label environmental sound classification task with $2000$ recordings of  $5$-second duration derived from $50$ environmental sound classes. The dataset is evaluated in a $5$-fold cross-validation setting.

    \item \textbf{IEMOCAP} \cite{busso2008iemocap}: A speech emotion recognition dataset with approximately 12 hours of audio data categorized into four emotion classes (happy, sad, angry, neutral). The emotion classification is evaluated using 5-fold cross validation.

    \item \textbf{US8K} \cite{salamon2014dataset}: A single-label audio scene classification dataset with 8,732 clips (less than 4 seconds) categorized into 10 urban sound classes. It is evaluated using 10-fold cross-validation.
    
    \item \textbf{SPCV2} \cite{warden2018speech}: A spoken command recognition dataset containing 84,843 training, 9,981 validation, and 11,005 test clips. The task is to classify each 1-second audio into one of 35 spoken commands.
    
    \item \textbf{VOX1} \cite{nagrani2017voxceleb}: A speaker identification task with 1,251 speakers. The dataset contains 138,361 training, 6,904 validation, and 8,251 test samples.
    
    \item \textbf{NSYNTH} \cite{engel2017neural}: A musical instrument classification task involving 4-second audio clips. The goal is to classify each clip into one of 11 instrument family classes.
\end{itemize}

\paragraph{Evaluation Metrics.} We adopt classification accuracy (Acc. \%) as the performance metric for all tasks.

For datasets with a validation set (SPCV2, VOX1, NSYNTH), we use the validation set for hyperparameter tuning and model selection, and report the final accuracy on the evaluation set. For ESC-50, US8K and IEMOCAP, we report the average accuracy across $5$, $10$ and $5$ fold validation respectively.
We finetune for $300$ epochs using ADAM optimizer, with cosine annealing of the learning rate down to \(10^{-6}\). The initial learning rate was found separately for each task using validation. 

\subsection{Downstream Evaluation and Comparison with Baselines}

Table~\ref{tab:downstream_results_ieee} presents the accuracy (\%) of different models evaluated on six diverse downstream tasks, all pretrained on a common dataset of (a) $200$ hours and (b) $2000$ hours.
All the models compared in this table use the same pre-training data and follow the pre-training recipes proposed in the respective repositories, which are available open-source. 
We compare: \textbf{(i)} SSAST model \cite{gong2022ssast}, \textbf{(ii)} HuBERT \cite{hsu2021hubert} which consisted of 2-stage pre-training, 
\textbf{(iii)} SSAST \cite{gong2022ssast} with masked-language modeling loss (replacing the reconstruction and the location identification loss proposed in the original framework) and, \textbf{(iv)} proposed \textbf{ULTRAS} framework, pretrained with joint spectro-temporal masking and predictive modeling on the same dataset. Our framework will also be made available upon paper acceptance. 

\begin{table*}[t!]
\caption{Impact (Acc. \%) of masking strategies and loss functions. Experiments are performed on 200 hr setup.}
\begin{center}
\resizebox{0.72\linewidth}{!}{ 
\begin{tabular}{|l|c|c|c||c|c|c|}
\midrule 
\multirow{2}{*}{\textbf{Model Variant}} & \multicolumn{3}{c||}{Audio} & \multicolumn{3}{c|}{Speech} \\ 
\cmidrule{2-7}
& \textbf{\textit{ESC-50}} & \textbf{\textit{US8k}} & \textbf{\textit{NSYNTH}} & \textbf{\textit{IEMOCAP}} & \textbf{\textit{VOX1}}  & \textbf{\textit{SPCV2}} \\
\midrule 
SSAST \cite{gong2022ssast} & 56.75 & 65.54 & 70.78 & 52.56 & 14.58 & 61.22 \\
-- + MLM & 77.65 & 78.46 & 75.32 & 60.49 & 32.19 & 78.67 \\
-- -- + Long-context masking & 82.20 & 79.31 & 75.52 & 61.06 & 36.33 & 82.17 \\
\system & \textbf{86.00} & \textbf{84.12} & \textbf{75.83} & \textbf{63.11} & \textbf{46.47} & \textbf{91.95} \\
\midrule
\end{tabular}
}
\end{center}
\label{tab:ablation_masking_loss}
\vspace{-0.2in}
\end{table*} 
The following are the key takeaways from this Table.
\begin{itemize}
    \item The baseline SSAST model is significantly inferior to the HuBERT setting on speech tasks. 
    \item The replacement of the SSAST loss function with the MLM loss leads to performance improvements across the board. 
    \item The proposed \system{} improves the SSAST+MLM setting on all the tasks considered here, highlighting the value of the joint spectral and temporal loss functions.
    \item All the systems show improvements from $200$ hour to $2000$ hour setup, illustrating the value of large-scale SSL pre-training for improved performance. 
    \item The proposed \system{} is observed to be superior in all tasks in 200 hour setting and it outperforms majority of the models in $2000$ hour setting. 
\end{itemize}

To evaluate the effectiveness of the proposal with state-of-art models, we compare its performance  with publicly available checkpoints of: HuBERT~\cite{hsu2021hubert}, SSAST~\cite{gong2022ssast}, BEATs~\cite{chen2022beats}, and EnCodecMAE~\cite{pepino2023encodecmae}. 
Note that, these pre-trained models use different amounts of data as well as mixtures of speech and audio corpora, while the proposed \system{} uses a relatively smaller dataset size of $2 000$ hours of data. 
All the downstream task evaluations pertain to the setup used in the previous setting, where the SSL model parameters are frozen during fine-tuning.

As shown in Table~\ref{tab:downstream_results_ieee_2000h},  
 the proposed model significantly improves over the SSAST setting on all tasks, while using a reduced pre-training dataset. The EnCodecMAE \cite{pepino2023encodecmae} illustrates the best performance on all the speech tasks, as it utilizes a significant amount of pre-training data ($11,000$ hours which includes about $6,000$ hours of speech data). 
 

\section{Ablation Studies}

\begin{table}[t!]
\caption{Performance (Acc. \%) for different $\lambda$ values in $200$ hr setup.}
\begin{center}
\resizebox{0.99\linewidth}{!}{ 
\begin{tabular}{|l|c|c|c|c|c|c|}
\midrule 
\midrule 
\multirow{2}{*}{$\boldsymbol{\lambda}$} & \multicolumn{3}{c|}{Audio} & \multicolumn{3}{c|}{Speech} \\ 
\cmidrule{2-7}
& \textbf{\textit{ESC}} & \textbf{\textit{US8k}} &  \textbf{\textit{NSYNTH}} & \textbf{\textit{IEMO}} & \textbf{\textit{VOX}}  & \textbf{\textit{SPCV2}} \\
\midrule 
0     & 82.20 & 79.31 & 75.52 & 61.06 & 36.33 & 82.17 \\
0.5   & 85.50 & 84.03 & 75.77 & 62.22 & 45.72 & 91.81 \\
\textbf{0.75} & \textbf{86.00} & \textbf{84.12} & \textbf{75.83} & \textbf{63.11} & \textbf{46.47} & \textbf{91.95} \\
0.99  & 84.10 & 83.18 & 74.20 & 62.25 & 45.33 & 89.76 \\
\midrule
\midrule 
\end{tabular}
}
\end{center}
\vspace{-0.1in} 
\label{tab:ablation_lambda}
\end{table}

\subsection{Long-Context Masking and Joint Loss}
In this analysis, we delve into the impact of the masked language modeling (MLM) loss as well the value of masking long-contextual windows ($160$ ms chunks). These results are shown in Table~\ref{tab:ablation_masking_loss}.
The SSAST \cite{gong2022ssast} framework is used as the starting point as it encodes spectral patches of the audio similar to the proposed \system. 
 
The introduction of the MLM loss improves the results on all tasks compared to the baseline SSAST setting. 
Further, the long-contextual masking, where all the spectral patches that belong to the same segment of $160$ ms duration are masked, further improves the performance on all tasks except for the emotion recognition task in IEMOCAP. 
The proposed \system{} framework modifies this setting using a joint spectro-temporal loss function.
This joint optimization significantly improves the performance on speech tasks 
while also marginally improving the performance on audio tasks. 
These experiments highlight the incremental value of the key steps in the \system{} framework, i.e., of masked language modeling loss functions on spectrogram patches, long-contextual masking and joint spectro-temporal loss functions. 
\subsection{ Effect of $\lambda$ in Joint Loss}
The loss function (Eq.~\ref{eqn:total-loss}) is a combination of spectral and temporal loss functions with a factor $\lambda$ regulating the weightage of the two losses. 
To study the impact of the  $\lambda$ used in our \system{} framework, we conduct an ablation where we set $\lambda$ as \{0, 0.5, 0.75, 0.99\}. A higher value of $\lambda$ increases the weight for the temporal loss. 
The best choice of $\lambda$  is $0.75$, as shown in   Table~\ref{tab:ablation_lambda}.  
We also observe that increasing $\lambda$ upto 0.75 leads to improved performance on speech tasks, while maintaining consistent performance on audio tasks.
\section{Conclusions}
\label{sec:ref}

In this work, we introduced \system, which proposed a self-supervised learning (SSL) framework for the joint modeling of speech and audio representations. 
 The proposal consists of encoding spectral patches belonging to relatively long-contextual windows ($160$ms).  
 The masked language modeling (MLM) setting is designed to  predict the temporal and spectral codes of the masked regions of the audio. This joint modeling is proposed as a way to force the encoding of spectral and temporal attributes of the acoustic signal, thereby enabling diverse downstream speech and audio tasks. 

 We perform experiments in a setting where the SSL framework is frozen while light-weight classification heads are learned on the downstream tasks. In this setting, we have compared the proposed \system{} with other frameworks using the same pre-training data ($200$ hour setup and $2000$ hour setup). When the pre-training data is matched, the proposed framework improves over prior works on audio tasks. The comparison of this work with other published models also reveals that the model achieves competitive performance while using significantly reduced training data size.
 
A set of ablation studies are conducted to understand the value of the masking strategy and the joint loss function. These experiments justify the design choices made in the proposed \system{} framework. 
These results highlight the importance of spectro-temporal representation learning and long-context masking in building general-purpose audio foundation models. Future directions include scaling ULTRAS to larger unlabeled corpora and extending it to tokenization schemes of audio large language models (LLMs).

\bibliographystyle{IEEEtran}
\bibliography{references}
\vspace{12pt}

\end{document}